\documentclass[twocolumn,showpacs,preprintnumbers,amsmath,amssymb,pra,aps,superscriptaddress]{revtex4-1}

\usepackage[pdftex]{graphicx}
\usepackage{amsmath}
\usepackage{color}
\usepackage{enumitem}

\newcommand{\ket}[1]{\left\lvert #1 \right\rangle}
\newcommand{\CZ}{\mathrm{CZ}}
\newcommand{\CZf}{\mathrm{CZ}_\mathrm{f}}
\newcommand{\CX}{\mathrm{CNOT}}
\newcommand{\ns}{\mathrm{ns}}
\newcommand{\Da}{D_\mathrm{a}}

\newcommand{\Di}{D_\mathrm{i}}
\newcommand{\Xa}{X_\mathrm{a}}

\newcommand{\Xd}{X_\mathrm{d}}
\newcommand{\Za}{Z_\mathrm{a}}

\newcommand{\Zd}{Z_\mathrm{d}}

\newcommand{\XL}{X_{\mathrm{L}}}

\newcommand{\ZL}{Z_{\mathrm{L}}}

\newcommand{\Fone}{f_1}
\newcommand{\Ftwo}{f_2}
\newcommand{\Fthree}{f_3}
\newcommand{\FtwoP}{f_2^{\mathrm{Park}}}
\newcommand{\FthreeP}{f_3^{\mathrm{Park}}}
\newcommand{\FoneI}{f_1^{\mathrm{Int}}}
\newcommand{\FtwoI}{f_2^{\mathrm{Int}}}

\newcommand{\Done}{D_1}
\newcommand{\Dtwo}{D_2}
\newcommand{\Dthree}{D_3}
\newcommand{\Dfour}{D_4}
\newcommand{\Xone}{X_1}
\newcommand{\Xtwo}{X_2}
\newcommand{\Zone}{Z_1}
\newcommand{\Ztwo}{Z_2}
\newcommand{\DF}{\Delta F}

\newcommand{\anh}{\alpha}
\newcommand{\MHz}{\mathrm{MHz}}
\newcommand{\GHz}{\mathrm{GHz}}

\newcommand{\toneq}{\tau_{\mathrm{1Q}}}
\newcommand{\ttwoq}{\tau_{\mathrm{2Q}}}

\makeatletter
   {\vskip\intextsep\parskip\z@
    \vbox\bgroup\centering\def\@captype{table}}%
   {\egroup\vskip\intextsep}
\makeatother

\begin{document}
\title{
Scalable quantum circuit and control for a superconducting surface code
}

\author{R.~Versluis}
\affiliation{Netherlands Organisation for Applied Scientific Research (TNO),  P.O. Box 155, 2600 AD Delft, The Netherlands}
\affiliation{QuTech, Delft University of Technology, P.O. Box 5046, 2600 GA Delft, The Netherlands}
\author{S.~Poletto}
\affiliation{QuTech, Delft University of Technology, P.O. Box 5046, 2600 GA Delft, The Netherlands}
\affiliation{Kavli Institute of Nanoscience, Delft University of Technology, P.O. Box 5046, 2600 GA Delft, The Netherlands}
\author{N.~Khammassi}
\affiliation{Computer Engineering, Delft University of Technology, Mekelweg 4, 2628 CD Delft, The Netherlands}
\author{N.~Haider}
\affiliation{Netherlands Organisation for Applied Scientific Research (TNO),  P.O. Box 155, 2600 AD Delft, The Netherlands}
\affiliation{QuTech, Delft University of Technology, P.O. Box 5046, 2600 GA Delft, The Netherlands}
\author{D.~J.~Michalak}
\affiliation{Components Research, Intel Corporation, 2501 NW 229th Ave, Hillsboro, OR 97124, USA}
\author{A.~Bruno}
\affiliation{QuTech, Delft University of Technology, P.O. Box 5046, 2600 GA Delft, The Netherlands}
\affiliation{Kavli Institute of Nanoscience, Delft University of Technology, P.O. Box 5046, 2600 GA Delft, The Netherlands}
\author{K.~Bertels}
\affiliation{Computer Engineering, Delft University of Technology, Mekelweg 4, 2628 CD Delft, The Netherlands}
\affiliation{Kavli Institute of Nanoscience, Delft University of Technology, P.O. Box 5046, 2600 GA Delft, The Netherlands}
\author{L.~DiCarlo}
\affiliation{QuTech, Delft University of Technology, P.O. Box 5046, 2600 GA Delft, The Netherlands}
\affiliation{Kavli Institute of Nanoscience, Delft University of Technology, P.O. Box 5046, 2600 GA Delft, The Netherlands}
\date{\today}

\begin{abstract}
We present a scalable scheme for executing the error-correction cycle of a monolithic surface-code fabric composed of fast-flux-tuneable transmon qubits with nearest-neighbor coupling. An eight-qubit unit cell forms the basis for repeating  both the quantum hardware and coherent control, enabling spatial multiplexing. This control  uses three fixed frequencies for all single-qubit gates and a unique frequency detuning pattern for each qubit in the cell. By pipelining the interaction and readout steps of ancilla-based $X$- and $Z$-type stabilizer measurements, we can engineer detuning patterns that avoid all second-order transmon-transmon interactions except those exploited in controlled-phase gates, regardless of fabric size. Our scheme is applicable to defect-based and planar logical qubits, including lattice surgery.
\end{abstract}
\maketitle

\tableofcontents

\section{Introduction}
\label{sec:introduction}
The scaling of small quantum processors~\cite{Kelly15, Corcoles15, Riste15, Debnath16, Cramer16} into large qubit arrays capable of fault-tolerant quantum computation (FTQC)~\cite{Nielsen00} is an outstanding challenge for leading experimental quantum information platforms~\cite{Martinis16, Brown16}. Modular~\cite{Nickerson14} and monolithic~\cite{Fowler12} approaches require a systems approach that simultaneously and compatibly addresses challenges in all layers of the quantum computer stack~\cite{Jones12}: from the quantum hardware at the low level, through classical control electronics in the middle, to software at the high level (i.e., micro-instruction sets, compilers, and high-level programming languages).

Currently, the surface code~\cite{Bravyi98,Fowler12, Terhal15} provides an experimentally attractive paradigm for FTQC owing to its modest requirements on the quantum hardware: only nearest-neighbor coupling is needed between qubits, and the error threshold falls robustly close to $1\%$ across a range of error models and error-decoding strategies, signficantly higher than those of Steane and Shor codes~\cite{Nielsen00}. In superconducting quantum integrated circuits based on circuit QED (cQED)~\cite{Blais04}, the error rate of single-qubit gates has reached $<0.1\%$~\cite{Barends14,Rol16}, while those of  two-qubit conditional-phase ($\CZ$) gates and measurement are $0.6\%$~\cite{Barends14} and $\sim 1\%$~\cite{Riste12b, Jeffrey14}, respectively.

The scalability of monolithic systems hinges on the ability to copy-paste a unit cell in the quantum plane, with suitable quantum interconnect between cells, and suitable classical interconnect to and from  the control plane. The latter pursuit is very active, with several groups developing
vertical (rather than the traditional lateral) interconnection of input/output (I/O) signals using through-the-wafer coaxial lines~\cite{Bruno16MM}, electro-mechanical sockets~\cite{Bejanin16}, and bump-bonding in flip-chip configuration~\cite{Rosenberg16MM}.

For true scalability, it is crucial that the unit cell also extend into the classical control plane. A unit cell for control signals opens the door to hardware simplification through spatial multiplexing, i.e., the selective routing of control signals (with minimal customization) to spatially separated components. While frequency-division multiplexing is already heavily exploited in cQED~\cite{Groen13, Jeffrey14, Riste15}, spatial multiplexing is in its infancy.  Precision control of same-frequency qubits using a microwave-frequency vector switch matrix (VSM) for pulse multicasting has only recently been demonstrated~\cite{Asaad16}.

In this paper, we propose a scalable scheme for the QEC cycle of a monolithic superconducting surface code by defining a concrete unit cell for both the quantum hardware and the control signals. We focus on a fabric of fast-flux-tunable transmon qubits interacting with nearest neighbors via flux-controlled conditional-phase $(\CZf)$ gates~\cite{Strauch03, DiCarlo09} realized by pulsing into the resonator-mediated $\ket{11} \leftrightarrow \ket{02}$ avoided crossing of the interacting transmon pair (numbers indicate excitation level). Our approach is compatible with adiabatic~\cite{DiCarlo09}, sudden~\cite{DiCarlo10} and fast-adiabatic~\cite{Martinis15, Barends14} use of these crossings.  Our eight-qubit unit cell uses three fixed frequencies for all single-qubit control and eight detuning sequences for two-qubit gates. This approach to classical control allows significant control hardware savings via spatial multiplexing. By pipelining the measurement of the two types of stabilizers of the surface code, we engineer detuning patterns avoiding all second-order transmon-transmon interactions except those exploited in $\CZf$ gates, regardless of fabric size. Our scheme allows changing the weight of stabilizer measurements by simple on/off masking of detuning pulses, making it applicable to both defect-based and planar logical qubits~\cite{Fowler12}, including lattice surgery~\cite{Horsman12}.

\section{Background}\label{sec:background}
\subsection{Surface code QEC cycle}

\begin{figure}
\includegraphics{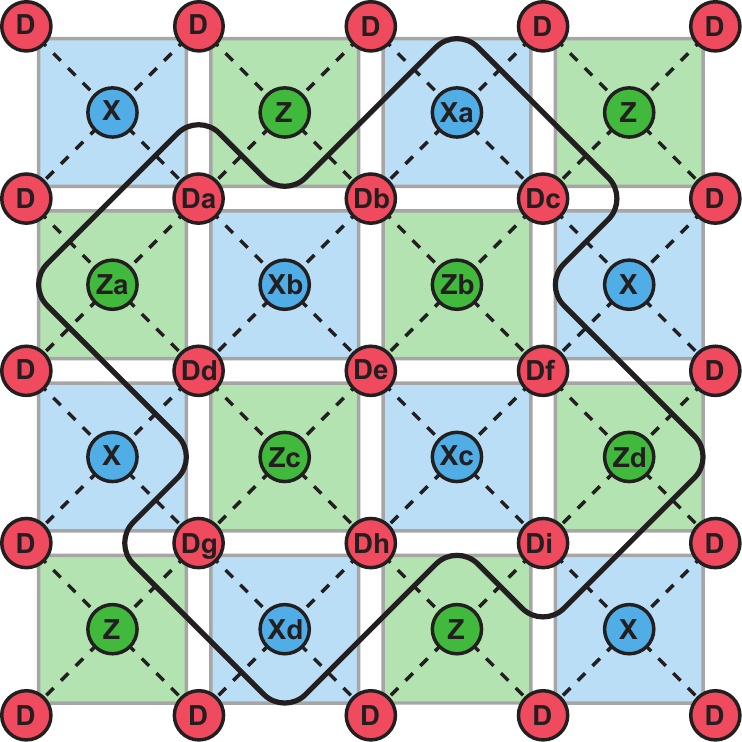}
\caption{\label{fig:Large_SC}
Layout of a surface-code fabric.  Red circles with $D$ labels represent data qubits. Blue (green) circles with $X$ ($Z$) labels represent ancillas performing $X$-type ($Z$-type) quantum parity checks of their nearest-neighbor data qubits. Each check is realized as an indirect quantum measurement, consisting of a coherent step involving pairwise interactions (dashed lines) followed by ancilla measurement. The delineated fabric of nine data qubits ($\Da$ through $\Di$) and eight ancillas ($\Xa$ through $\Xd$ and $\Za$ through $\Zd$) constitutes the distance-3 planar logical qubit named Surface-17.}
\end{figure}

\begin{figure}
	\includegraphics{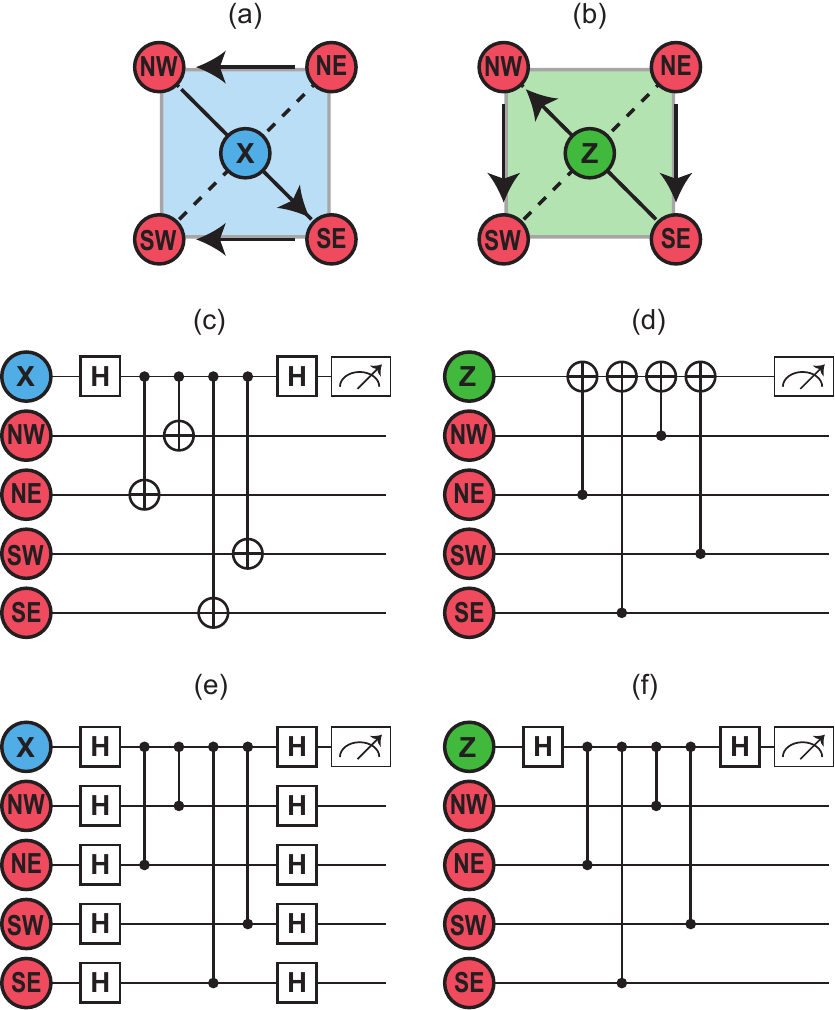}
	\caption{\label{fig:Plaquettes}
$X$-type (a) and $Z$-type (b) plaquettes. Data qubits are labelled according to their position relative to the ancilla (NE=northeast, NW=northwest, SE=southeast, and SW=southwest). Standard circuits for measuring $X$-type (c, e) and $Z$-type (d, f) stabilizers indirectly using ancillas, using  $\CX$ (c, d) or $\CZ$ (e, f) as the primitive data-ancilla interaction. The order of two-qubit gates, NE-NW-SE-SW (NE-SE-NW-SW) for $X$-type ($Z$-type), ensures that all data qubits common to  adjacent plaquettes do their interactions with one ancilla before the other, and also provides resilience to ancilla errors in Surface-17~\cite{Tomita14}. Using the relations $H=Y_{+90}Z=ZY_{-90}$, one can see that the opening and closing $H$ gates can be replaced by $Y_{-90}$ and $Y_{+90}$ rotations, respectively.}
\end{figure}

A surface-code fabric consists of the two-dimensional square lattice of data-carrying qubits shown in Fig.~\ref{fig:Large_SC}. The stabilizers of this code are the $X$-type ($Z$-type) parity operators $\prod_i X_i$ $\left(\prod_i Z_i \right)$, where $i$ denotes data qubits on the corners of the blue (green) plaquettes. Conventionally, these stabilizers are measured indirectly using ancilla qubits positioned at the center of the plaquettes, forming a second square lattice. Standard circuits for measuring $X$- and $Z$-type stabilizers, shown in Fig.~\ref{fig:Plaquettes}, combine a sequence of coherent interactions of the ancilla with its nearest-neighbor data qubits, followed by projective ancilla measurement.

Using controlled-not ($\CX$) gates as the fundamental interaction,  $X$-type and $Z$-type stabilizer measurements can be fully parallelized with circuit depth seven. We define circuit depth as the number of operations on each ancilla per QEC cycle, counting in measurement but excluding ancilla initialization [we assume Pauli frame updating (PFU)~\cite{Knill05, Terhal15} is used for data and ancilla qubits]. The order of two-qubit gates in Fig.~\ref{fig:Plaquettes} is important for two reasons~\cite{Tomita14}. First,
data qubits common to adjacent plaquettes must do all their interactions with one ancilla before the other. Second, the S (N) pattern for $X$-type ($Z$-type) stabilizers provides resilience to single ancilla-qubit errors even in small distance-three surface codes such as Surface-17. This circuit consists of the patch delineated in Fig.~\ref{fig:Large_SC}, with nine data qubits (labelled $\Da$ to $\Di$), four ancillas ($\Xa$ to $\Xd$) for $X$-type stabilizer measurements, and four ancillas
($\Za$ to $\Zd$) for $Z$-type stabilizer measurements.

\begin{figure}
	\includegraphics{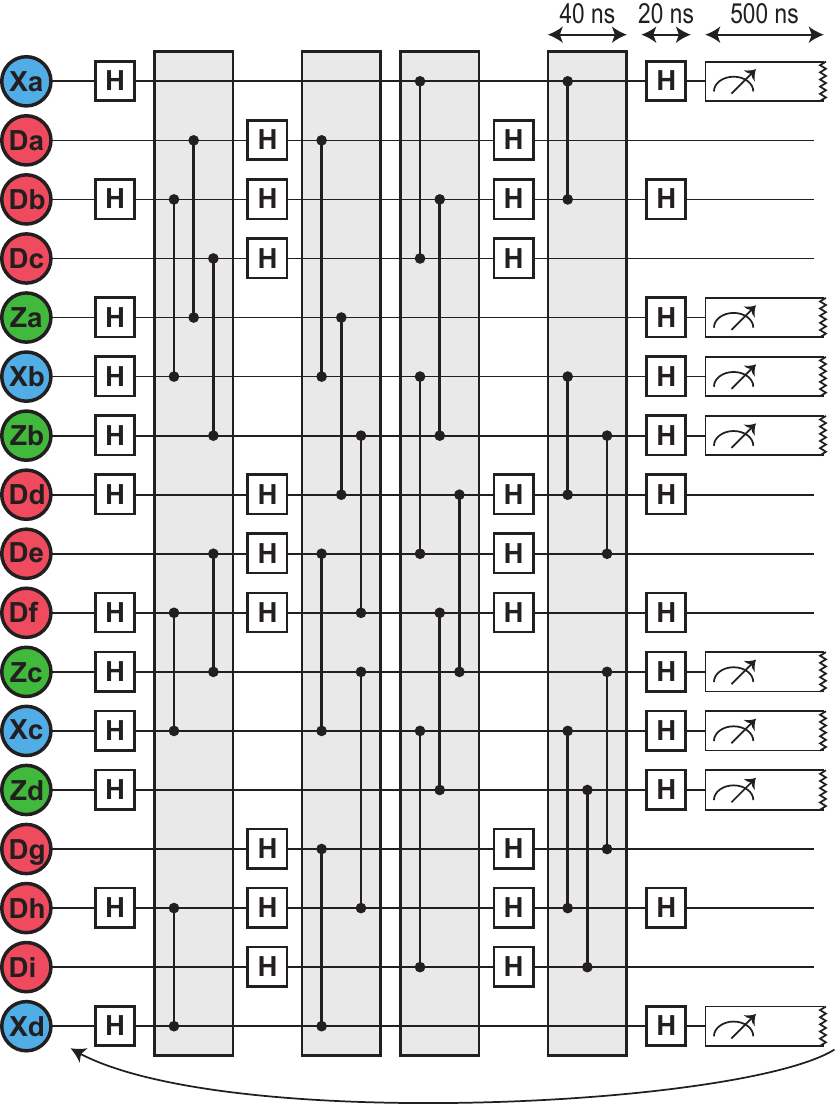}
	\caption{\label{fig:Parallelized} Depth-nine quantum circuit for parallelized $X$- and $Z$-stabilizer measurements in Surface-17 using $\CZ$ gates. The six $\CZ$ gates inside each gray box are executed simultaneously. Typical values  of gate and readout times are indicated at the top. The bottom arrow represents the looping of QEC cycles. Qubits are labelled as in Fig.~\ref{fig:Large_SC}.}
\end{figure}

When the two-qubit gate is $\CZ$, parallelizing the stabilizer measurements of Surface-17 requires depth nine because of non-commutation between Hadamard $(H)$ gates and $\CZ$ gates. The full circuit for the parallelized QEC cycle of Surface-17 using $\CZ$ gates is shown in Fig.~\ref{fig:Parallelized}. Using gate and measurement times from recent experiments  ($\toneq=20~\ns$ for single-qubit gates, $\ttwoq=40~\ns$ for $\CZf$ gates, and $500~\ns$ for ancilla readout and photon depletion in readout resonator), the QEC cycle will complete in $740~\ns$.

\subsection{Limitations of fully parallelized $X$- and $Z$-type stabilizer measurements using $\CZf$ gates}
\label{sec:parallel_execution}
On paper, it is straightforward to compose a depth-nine quantum circuit for the fully parallelized QEC cycle of a surface-code fabric of arbitrary size. However, to the best of our knowledge following numerous failed  attempts, the full parallelization of $X$- and $Z$-type stabilizer measurements makes it impossible to realize a scalable implementation with $\CZf$ gates that satisfies all of the following desirable properties:
\begin{itemize}
\item
Microwave pulses for single-qubit gates should be applied at a fixed, small number of frequencies.
\item
Transmons should maximally exploit their coherence sweetspot~\cite{Schreier08}.
\item
Flux-pulsed transmons should not cross any other interaction zones on their way to or from the intended
$\ket{11} \leftrightarrow \ket{02}$ avoided crossings realizing the $\CZf$ gate.
\item
The flux-pulsing schemes should be extensible to a surface code of arbitrary size using a fixed number of detuning sequences and a fixed detuning range.
\item
The implementation should be compatible with logical qubit operations.
\end{itemize}

We have found frequency arrangements and flux-pulse sequences that meet the first three criteria.
However, all of these solutions require a growing number of detuning sequences and detuning ranges as the fabric expands, in order to avert all other interactions on the way to and from the $\ket{11} \leftrightarrow \ket{02}$ avoided crossings of $\CZf$ gates. Furthermore, these solutions seem practically infeasible already for distance five (Surface-49~\cite{Horsman12}). To our knowledge, no fully parallel solution exists with a fixed number of detuning sequences and a fixed detuning range. In the next section, we introduce a pipelined (rather than parallelized) version of the QEC cycle that simultaneously meets the five desirable properties for a fabric of arbitrary size.

\section{The pipelined QEC cycle}
\label{sec:implementation}

Our scalable scheme, which we term \textit{pipelined QEC cycle}, combines four key elements:

\begin{enumerate}
	\item[A.] Repeating unit cells of eight qubits;
	\item[B.] Pipelined $X$- and $Z$-type stabilizer measurements;
	\item[C.] Three frequencies for single-qubit control;
    \item[D.] Eight detuning sequences implementing the requisite $\CZf$ gates, realizable by on/off masking of three flux-pulse primitives.
\end{enumerate}
We now introduce these elements in detail.

\subsection{Unit cell}
\begin{figure}
	\includegraphics{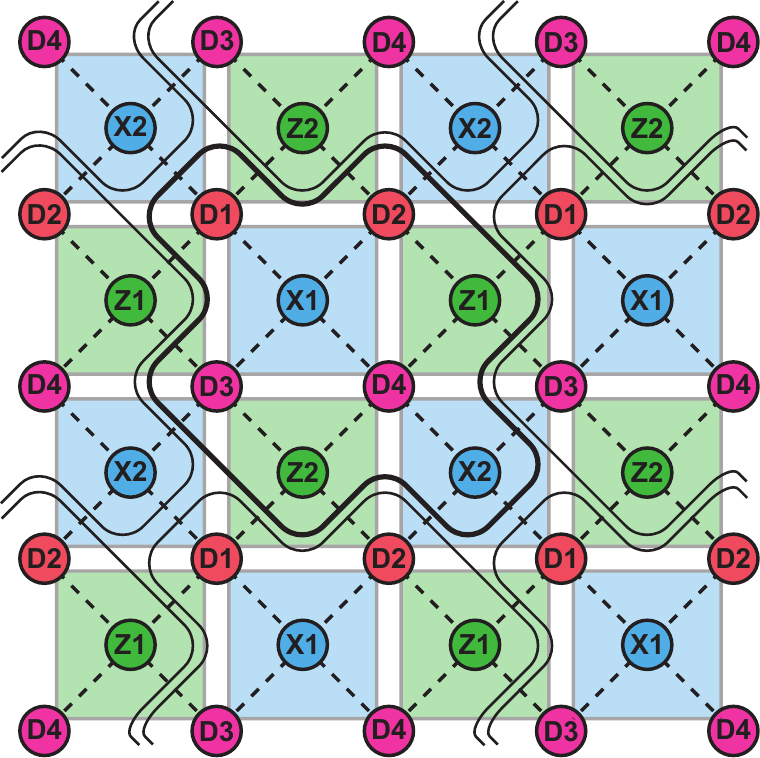}
	\caption{\label{fig:SC_layout} Composing the surface-code fabric by repetition of 8-qubit unit cells. Red and pink circles represent data qubits,  blue (green) circles represent ancillas for $X$-type ($Z$-type) stabilizer measurements, and dashed lines represent nearest-neighbor couplings. Dot colors also indicate the frequency for single-qubit microwave control (red for $\Fone$, greed and blue for $\Ftwo$, and pink for $\Fthree$). Contours delineate unit cells (with qubits named $\Done$ to $\Dfour$, $\Xone$, $\Zone$, $\Xtwo$, and $\Ztwo$).}
\end{figure}

The first element is a unit cell (Fig.~\ref{fig:SC_layout}) from which a surface code of arbitrary size can be assembled by repetition (and truncation at boundaries). A unit cell contains four data qubits ($\Done$ to $\Dfour$) and four ancillas ($\Xone$, $\Xtwo$, $\Zone$, and $\Ztwo$). Crucially, the cell is the fundamental unit of repetition not just for the quantum hardware. It is also the unit of repetition for all coherent control.

\subsection{Pipelining of $X$-type and $Z$-type stabilizer measurements}
\begin{figure*}
	\includegraphics{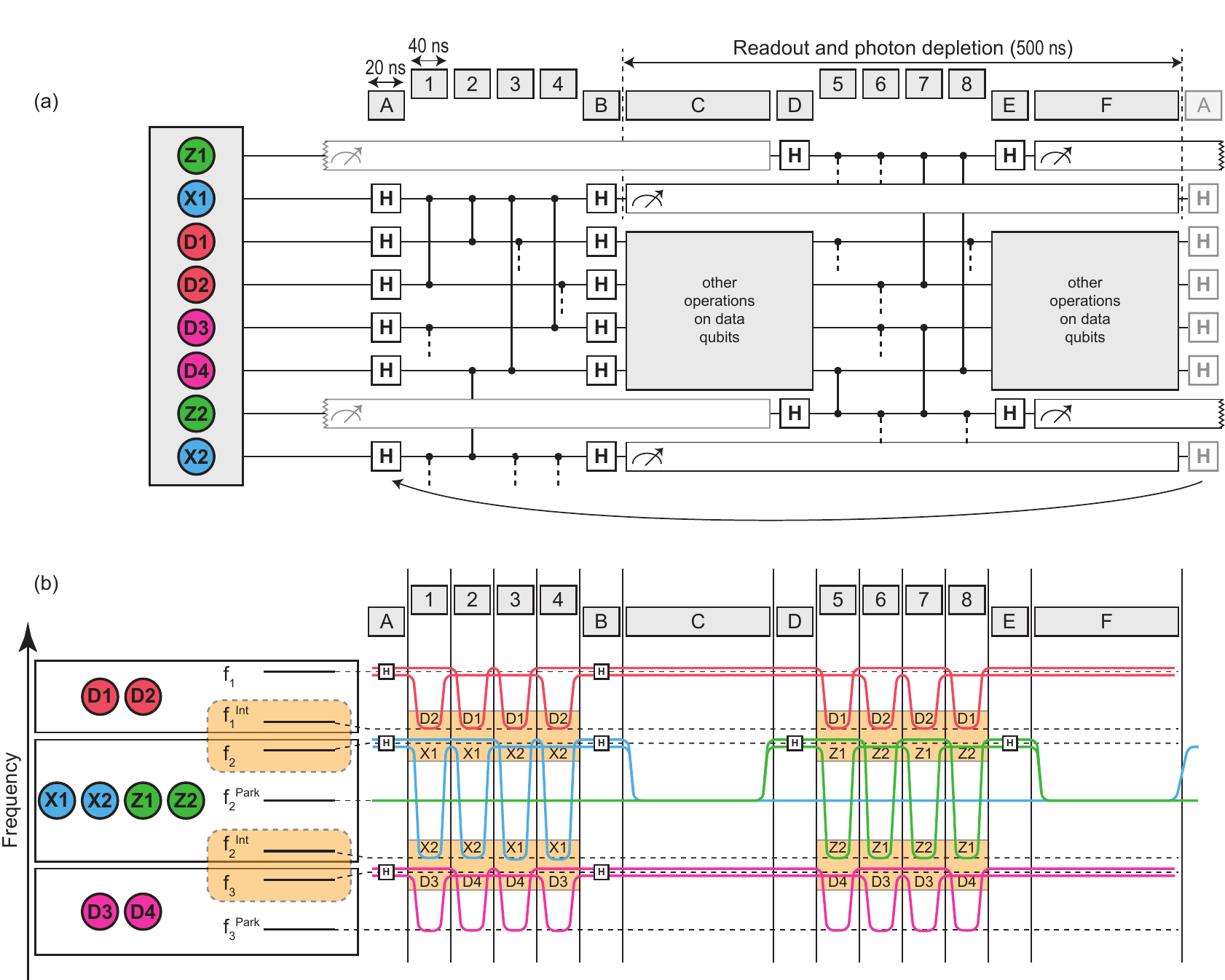}
	\caption{\label{fig:Pipelined} (a) Unit cell quantum circuit for pipelined $X$- and $Z$-type stabilizer measurements. Qubits are labelled as in Fig.~\ref{fig:SC_layout}. Time slots 1 to 4 (5 to 8) are for $X$-type ($Z$-type) stabilizer $\CZ$ gates. Time slots A and B (D and E) are for $X$-type ($Z$-type) stabilizer single-qubit gates. Time slots C and F allow (optional) operations on data qubits while $X$-ancillas ($Z$-ancillas) are measured during time slots D through F (time slots A through C). The bottom arrow indicates the repetition of QEC cycles. Elements in transparent gray belong to the previous or next QEC cycle. (b) Frequency arrangement and detuning sequences for qubits in the unit cell: single-qubit gates on $\Done$ and $\Dtwo$ ($\Dthree$ and $\Dfour$) are performed at $\Fone$ ($\Fthree$), while those on ancillas are performed at $\Ftwo$. Ancilla measurements are performed at $\FtwoP$. $\CZf$ gates are performed between qubits at $\FoneI$ and $\Ftwo$ or at $\FtwoI$ and $\Fthree$. No interactions take place at $\Fone$ or the parking frequencies $\FtwoP$ and $\FthreeP$. Small offsets are added to some detuning sequences to clarify the distinction between sequences for $\Done$ and $\Dtwo$, $\Xone$ and $\Xtwo$, $\Zone$ and $\Ztwo$, and $\Dthree$ and $\Dfour$.}
\end{figure*}

The second element is the pipelined execution of the $X$- and $Z$-type stabilizer measurements. The pipelining concept is illustrated in Fig.~\ref{fig:Pipelined}(a). While stabilizer measurements of one type always run simultaneously, the coherent and readout steps of ancillas of the other type are interleaved. In other words, ancillas of one type undergo coherent steps while ancillas of the other type are measured. Time slots A and B (D and E) are for single-qubit gates pertinent to the  $X$-type ($Z$-type) stabilizer measurements, while slots 1 to 4 (5 to 8) are for two-qubit gates. Note that nine of the $\CZ$ gates involve two qubits within the cell, while fourteen involve one qubit from a neighboring unit cell.

Generally, ancilla measurement (including any photon depletion of the readout resonator) will take longer than the coherent steps, leaving
time to perform operations on the data qubits in steps C and F while all ancillas are measured. Possible operations include logical gate operations, refocusing pulses, or single-qubit gates performing error correction. Clearly, performing such operations during steps C or F would not increase the QEC cycle time.

Pipelining offers several advantages. First, it compresses the stabilizer measurements to depth seven, two single-qubit-gate steps less than fully-parallelized quantum circuits (such as Fig.~\ref{fig:Parallelized} for Surface-17). A second and more crucial advantage is the ability to scale without increasing the number of frequencies for single-qubit control or qubit detuning sequences, as explained next.

\subsection{Single-qubit control and detuning sequences}
The third and fourth elements are best described together. Figure~\ref{fig:Pipelined}(b) presents our choice of frequencies for single-qubit control and the qubit-specific detuning sequences for realizing the two-qubit QEC cycle interactions. Single-qubit gates on data qubits (steps A, B, D, and E) are performed at frequencies $\Fone$ and $\Fthree$ (alternating in data-qubit rows), while those on ancillas are performed at intermediate frequency $\Ftwo$. Note that with only nearest-neighbor coupling, two distinct frequencies (one for ancilla qubits and one for data qubits) reduce the exchange coupling between same-frequency qubits to fourth order (qubit-resonators, resonator-qubit, qubit-resonator, resonator-qubit). When extending to the proposed three frequencies, this also allows engineering the detuning sequences so that no transmon crosses any other second-order interaction zone on the way to or from the $\ket{11} \leftrightarrow \ket{02}$ avoided crossings exploited in the $\CZf$ gates.

During steps 1-4 and 5-8, transmons are flux pulsed to a discrete set of frequencies, depending on whether they interact, idle, or are measured: $\Done$ and $\Dtwo$ to $\Fone$ or $\FoneI$; ancillas to $\Ftwo$, $\FtwoP$ or $\FtwoI$; and $\Dthree$ and $\Dfour$ to $\Fthree$ or $\FthreeP$. $\CZ$ gates occur between transmons at $\FoneI \approx \Ftwo-\anh$ and $\Ftwo$, and between transmons at $\FtwoI \approx \Fthree-\anh$ and $\Fthree$. Here, $\anh \sim -300~\MHz$ is the transmon anharmonicity~\cite{Koch07}. Transmons at $\Fone$, $\FtwoP$, and $\FthreeP$ are hidden away and do not interact.

\subsection{Constructing detuning sequences by masking of primitive flux pulses}
 The frequency detuning patterns during interaction steps 1 through 4 and 5 through 8 can be synthesized by on/off masking of three flux-pulse primitives using a switch matrix:
 A first primitive detuning data qubits of type $\Done$ and $\Dtwo$ from $\Fone$ to $\FoneI$, a second one detuning ancillas from $\Ftwo$ to $\FtwoI$, and a third one detuning data qubits of type  $\Dthree$ and $\Dfour$ from $\Fthree$ to $\FthreeP$. For example, the detuning sequence for $\Dtwo$ in Fig.~\ref{fig:Pipelined}(b) can be synthesized by masking the pulse primitive on (off) at steps 1, 4, 6, and 7 (2, 3, 5, and 8).

\section{Compatibility with logical qubit operations}
\label{sec:Defect}
Two types of logical qubits can be envisioned for surface code: defect-based~\cite{Fowler12} and planar~\cite{Horsman12}. Defect-based logical qubits are introduced by stopping the measurement of one or two stabilizers ($X$-type for \textit{rough} logical qubits, and $Z$-type for \textit{smooth} ones~\cite{Fowler12}). In our scheme, turning stabilizer measurements fully off can be accomplished in either of two ways. One is to mask off the $H$ gates of the corresponding ancilla using the VSM, without changing the detuning sequence or stopping the ancilla measurement. If the ancilla is in $\ket{0}$, all its $\CZf$ gates are inactive and there is no net action on the logical qubit. If it starts in $\ket{1}$, the stabilizer operator (not its measurement) is applied. This performs a logical $\XL$ $(\ZL)$ gate on a rough (smooth) qubit. The ancilla measurements provide the key input allowing the decoder to keep track of the action by PFU. A second way to turn a stabilizer fully off is to mask off all the flux-pulse primitives in the interaction step.

Logical operations, such as move and braiding operations on defect-based qubits~\cite{Fowler12}, and lattice surgery on planar ones~\cite{Horsman12}, also require dynamically changing the weight of specific stabilizer measurements, i.e., selectively removing specific data qubits from the quantum parity checks. In our scheme, this can easily be achieved by selective on/off masking of flux-pulse primitives. For example, removing a qubit of type $\Dtwo$ from the $X$-type stabilizer measurement below it simply requires masking off the pulse primitive at step 1. The order of the two-qubit gates can also be changed by masking.

\section{Implementation details and variations}
\label{sec:implementation details}

\subsection{Choosing the frequencies}

Ideally, $\Fone$ ($\Fthree$) would match the sweetspot frequency of $\Done$ and $\Dtwo$ ($\Dthree$ and $\Dfour$), and $\Ftwo$ would match that of all ancillas, to minimize dephasing from $1/f$ flux noise. In practice, $\Fone$ would be set to the lowest sweetspot frequency among all transmons labelled $\Done$ or $\Dtwo$, and so on. It is straightforward to expand the circuit of Fig.~\ref{fig:Pipelined}(a) with refocusing single-qubit gates to minimize dephasing in the transmons that are not at their sweetspot during single-qubit control~\cite{OBrien16}.

The frequencies $\Fone$, $\Ftwo$, $\Fthree$, $\FtwoP$, and $\FthreeP$ must be chosen so that residual interactions during single-qubit gates can be tolerated.
For simplicity, we consider a uniform detuning scale $\DF = \Fone-\FoneI=\Ftwo-\FtwoP=\FtwoP-\FtwoI=\Fthree-\FthreeP$. The single-qubit gate error is
$\varepsilon \sim \left(\frac{\xi^2}{4\pi\DF}\toneq\right)^2$, where $\xi$ is the $\ket{11} \leftrightarrow \ket{02}$ coupling strength~\cite{DiCarlo09}. For fast-adiabatic $\CZf$ gates, $\ttwoq\approx 2\frac{\pi}{\xi}$. Thus, $\varepsilon\sim~10^{-2}\ (10^{-3})$ for $\toneq=20~\ns$ and $\ttwoq=40~\ns$ requires $\DF \sim 400~\MHz\ (1.2~\GHz)$.

\subsection{Frequency arrangement variations}

\begin{figure}
	\includegraphics{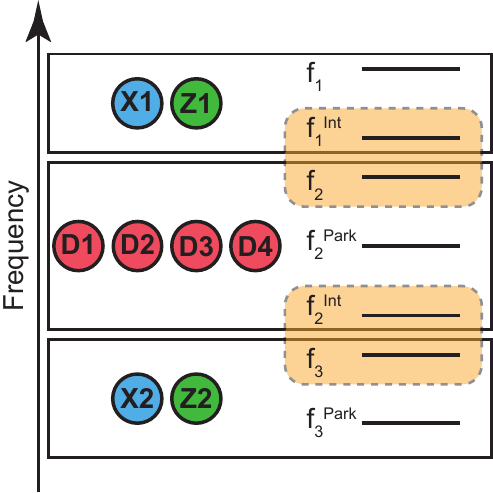}
	\caption{\label{fig:Ancilla_out} Alternative frequency arrangement for qubits in the unit cell, with ancillas $\Xone$ and $\Zone$ ($\Xtwo$ and $\Ztwo$) at the outer frequency $\Fone$ ($\Fthree$) and data qubits at the inner frequency $\Ftwo$.}
\end{figure}

There exist other possible frequency arrangements than that shown in Fig.~\ref{fig:Pipelined}(b). For example, consider the inverted arrangement with all
data qubits at $\Ftwo$ and the ancillas at the outer frequencies. Figure~\ref{fig:Ancilla_out} shows one of these configurations, with $\Xone$ and $\Zone$ ($\Xtwo$ and $\Ztwo$) at $\Fone$ ($\Fthree$). It is straightforward to modify the detuning sequences for this arrangement to also avert all unwanted interactions. However, upon comparing this alternative to the original arrangement, we observe a key difference making the original preferable for a cQED implementation with flux-tuneable transmons. Specifically, the original exactly balances the number of interaction steps in which qubits can remain at their upper frequency (i.e., at or closest to their coherence sweetspot), while the flipped arrangement allows this on just two (out of eight) steps for data qubits and zero or four (out of four) steps for ancilla qubits. The reduced data-qubit dephasing during the coherent steps will lead to a lower logical error rate. Note that this advantage of the original arrangement is made possible by lowering the ancillas to $\FtwoP$ for their measurement, at which the additional dephasing is innocuous in view of the measurement-induced projection.

Residual single-qubit gate cross-talk between $\Done$ and $\Dtwo$ ($\Dthree$ and $\Dfour$) can be reduced by breaking the degeneracy in frequency $\Fone$ ($\Fthree$), which requires increasing the number of primitive pulses from three to five, or even in $\Ftwo$, further increasing the number of primitive pulses to eight.

\subsection{Switch matrix}
 A digitally addressable switch matrix and primitive-flux-pulse-synthesizer operating at room temperature (cryogenically) are exciting challenges for the near (long) term. The switch matrix should allow qubit-specific customization of the flux-pulse primitives, including fine adjustment of delay (to compensate cable-length mismatch), amplitude (for fine tuning of two-qubit and single-qubit phase), and dc offset (for tuning to $\Fone$, $\Ftwo$ and $\Fthree$).

\section{Conclusion and outlook}

\begin{figure}
  \centering
  \includegraphics[width=\columnwidth]{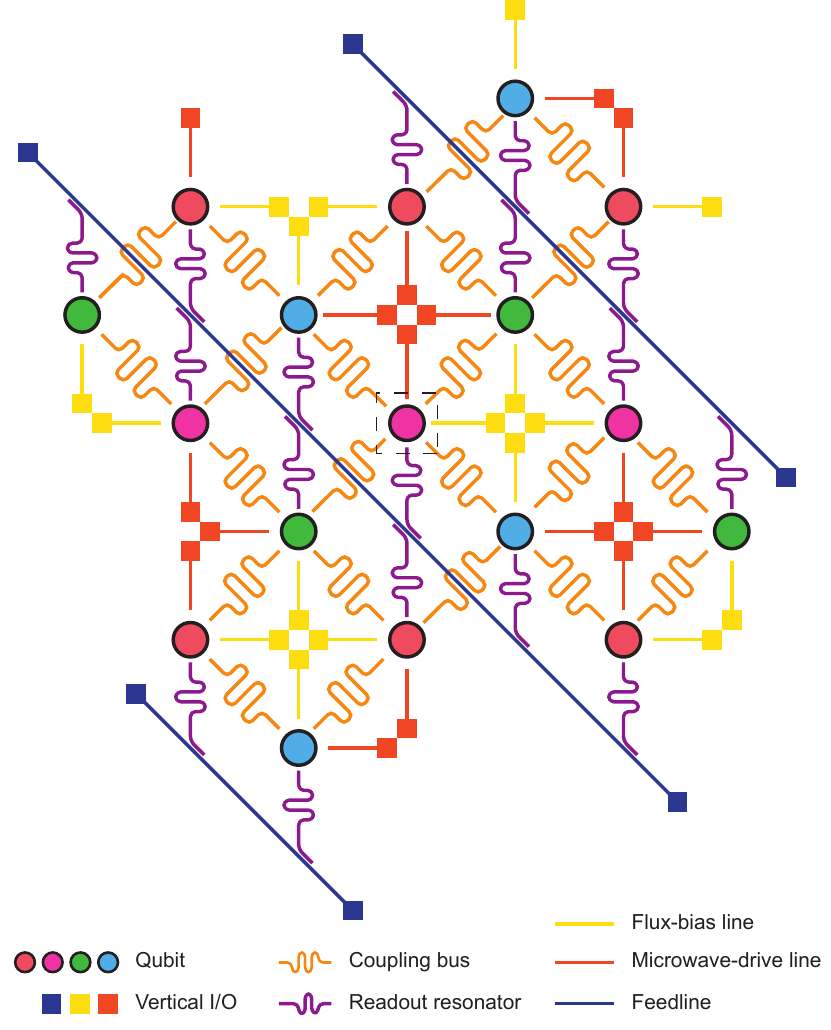}
  \caption{Schematic of the targeted realization of Surface-17 in a planar cQED architecture with vertical I/O. Every transmon (represented by a circle) has dedicated flux control line, microwave-drive line, and readout resonator. Dedicated bus resonators mediate interactions between nearest-neighbor data and ancilla qubits. Readout resonators are simultaneously interrogated using frequency-division multiplexing in diagonally-running feedlines. The central dashed square corresponds to the scanning electron micrograph in Fig.~\ref{fig:Implementation}(a).  Note that a similar cQED chip design is envisioned in Ref.~\cite{Bejanin16}. Our scheme reduces the number of feedlines by bridging these over bus resonators.}
  \label{fig:CQEDS17}
\end{figure}

\begin{figure}
  \centering
  \includegraphics[width=0.8\columnwidth]{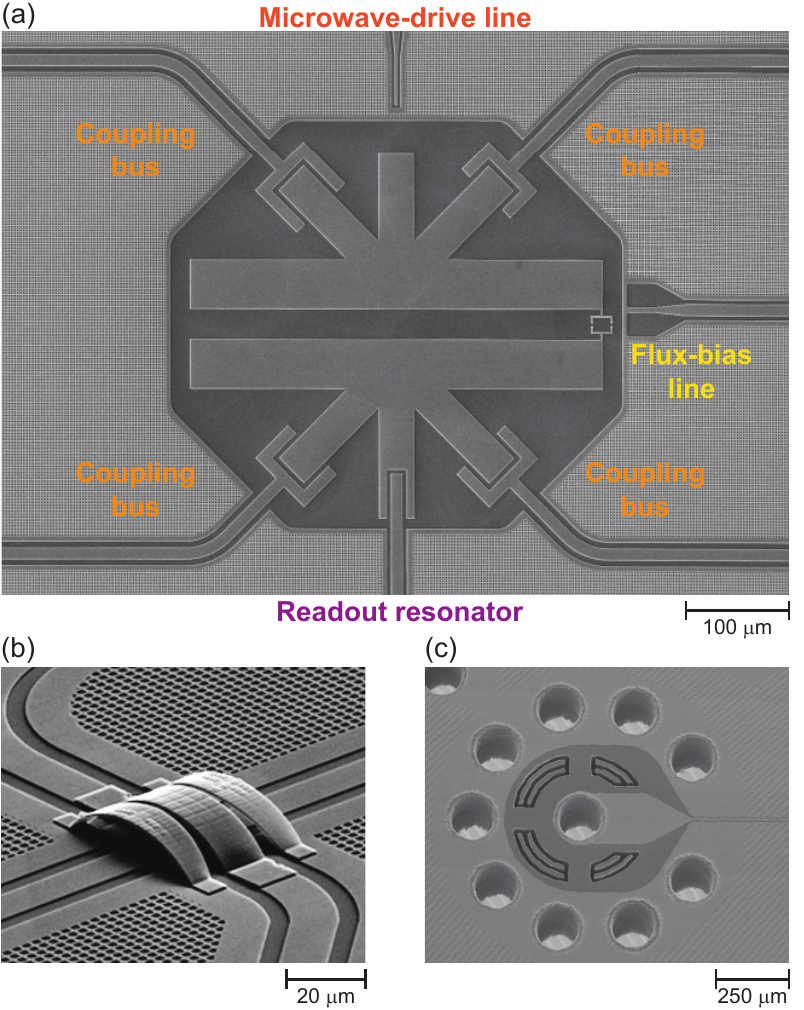}
  \caption{Scanning electron micrographs of the key components of our cQED implementation of surface-code fabric: (a) starmon~\cite{Poletto17}, (b) transmission-line crossover~\cite{Riste15}, and (c) vertical I/O~\cite{Bruno16MM}. (a) Starmon is a seven-port transmon connecting to four bus resonators (corners), one readout resonator (bottom), one microwave-drive line (top), and one flux-bias line (right). The ports in this image correspond directly to those in the dashed square in Fig.~\ref{fig:CQEDS17}. (b) Transmission-line crossovers bridge readout feedlines over bus resonators. (c) Vertical coaxial lines couple all input and output signals.}
  \label{fig:Implementation}
\end{figure}

We have presented a concrete scheme for the QEC cycle of an arbitary-size surface code implemented with flux-tuneable transmons. The scheme combines four key concepts: an eight-qubit unit cell as the basis  for repetition of  quantum hardware and control signals; pipelining of $X$- and $Z$-type stabilizer measurements; a fixed set of three frequencies for single-qubit control; and a fixed set of eight detuning sequences implementing the requisite controlled-phase gates. These eight detuning sequences can be composed by on-off masking of three flux-pulse primitives.

Experimental efforts are underway to implement and evaluate the pipelined QEC cycle in Surface-17. We pursue a realization of the Surface-17 quantum hardware with transmons in a planar cQED architecture~\cite{Blais04} made extensible using vertical interconnect for all input and output signals (Fig.~\ref{fig:CQEDS17}).  Each transmon will have a dedicated flux line~\cite{DiCarlo09} allowing control of its transition frequency on nanosecond timescales, a dedicated microwave-drive line, and a dedicated dispersively-coupled resonator for readout. We opt for coupling nearest-neighbor data and ancilla transmons via bus resonators (rather than direct capacitance~\cite{Kelly15}). Figure~\ref{fig:Implementation}(a) shows a prototype seven-port transmon developed in our lab, nicknamed \textit{starmon}~\cite{Poletto17}. The vertical I/O will be realised either using through-silicon-vias [Fig.~\ref{fig:Implementation}(c)] or bump bonding in a flip-chip arrangement.

Experimental efforts are also underway to demonstrate the scalability of the classical control plane. Diagonally running feedlines coupled to readout resonators will allow simultaneous readout by frequency-division multiplexing~\cite{Groen13,Riste15}, reducing the need for cryogenic amplifiers and microwave electronics (circulators, etc.), as well as homodyne detection systems at room temperature. While frequency multiplexing for readout is common in cQED, the simultaneous readout of nine qubits using one feedline as required by Surface-17 is not yet achieved. Finally, demonstrating the control-hardware savings achievable by spatial multiplexing is an immediate priority. Single-qubit control at $\Fone$, $\Ftwo$ and $\Fthree$ will make use of a next-generation VSM (follow-up to~\cite{Asaad16}) to independently route precision DRAG~\cite{Motzoi09,Chow10b} pulses to same-frequency qubits, with significant savings in microwave sources. Finally, a switch matrix for constructing frequency detuning sequences by on/off masking of flux-pulse primitives is envisioned for room temperature. A cryogenic implementation remains highly attractive for the longer term.

\section*{Acknowledgements}
We thank C.~C. Bultink, M.~A.~Rol, X.~Fu, C.~Garc\'{i}a-Almudever, L.~Riesebos, D.~Deurloo, B.~Criger, T.~E.~O'Brien, and B.~Tarasinski for helpful discussions.
This research is supported by the Dutch Organisation for Scientific Research on Matter (FOM), Intel Corporation, and by the Office of the Director of National Intelligence (ODNI), Intelligence Advanced Research Projects Activity (IARPA), via the U.S. Army Research Office grant W911NF-16-1-0071. The views and conclusions contained herein are those of the authors and should not be interpreted as necessarily representing the official policies or endorsements, either expressed or implied, of the ODNI, IARPA, or the U.S. Government. The U.S. Government is authorized to reproduce and distribute reprints for Governmental purposes notwithstanding any copyright annotation thereon.

%\bibliographystyle{apsrev4-1}
%\bibliography{../Paper_resources/References/References_cQED}

\begin{thebibliography}{36}%
\makeatletter
\providecommand \@ifxundefined [1]{%
 \@ifx{#1\undefined}
}%
\providecommand \@ifnum [1]{%
 \ifnum #1\expandafter \@firstoftwo
 \else \expandafter \@secondoftwo
 \fi
}%
\providecommand \@ifx [1]{%
 \ifx #1\expandafter \@firstoftwo
 \else \expandafter \@secondoftwo
 \fi
}%
\providecommand \natexlab [1]{#1}%
\providecommand \enquote  [1]{``#1''}%
\providecommand \bibnamefont  [1]{#1}%
\providecommand \bibfnamefont [1]{#1}%
\providecommand \citenamefont [1]{#1}%
\providecommand \href@noop [0]{\@secondoftwo}%
\providecommand \href [0]{\begingroup \@sanitize@url \@href}%
\providecommand \@href[1]{\@@startlink{#1}\@@href}%
\providecommand \@@href[1]{\endgroup#1\@@endlink}%
\providecommand \@sanitize@url [0]{\catcode `\\12\catcode `\$12\catcode
  `\&12\catcode `\#12\catcode `\^12\catcode `\_12\catcode `\%12\relax}%
\providecommand \@@startlink[1]{}%
\providecommand \@@endlink[0]{}%
\providecommand \url  [0]{\begingroup\@sanitize@url \@url }%
\providecommand \@url [1]{\endgroup\@href {#1}{\urlprefix }}%
\providecommand \urlprefix  [0]{URL }%
\providecommand \Eprint [0]{\href }%
\providecommand \doibase [0]{http://dx.doi.org/}%
\providecommand \selectlanguage [0]{\@gobble}%
\providecommand \bibinfo  [0]{\@secondoftwo}%
\providecommand \bibfield  [0]{\@secondoftwo}%
\providecommand \translation [1]{[#1]}%
\providecommand \BibitemOpen [0]{}%
\providecommand \bibitemStop [0]{}%
\providecommand \bibitemNoStop [0]{.\EOS\space}%
\providecommand \EOS [0]{\spacefactor3000\relax}%
\providecommand \BibitemShut  [1]{\csname bibitem#1\endcsname}%
\let\auto@bib@innerbib\@empty
%</preamble>
\bibitem [{\citenamefont {Kelly}\ \emph {et~al.}(2015)\citenamefont {Kelly},
  \citenamefont {Barends}, \citenamefont {Fowler}, \citenamefont {Megrant},
  \citenamefont {Jeffrey}, \citenamefont {White}, \citenamefont {Sank},
  \citenamefont {Mutus}, \citenamefont {Campbell}, \citenamefont {Chen} \emph
  {et~al.}}]{Kelly15}%
  \BibitemOpen
  \bibfield  {author} {\bibinfo {author} {\bibfnamefont {J.}~\bibnamefont
  {Kelly}}, \bibinfo {author} {\bibfnamefont {R.}~\bibnamefont {Barends}},
  \bibinfo {author} {\bibfnamefont {A.}~\bibnamefont {Fowler}}, \bibinfo
  {author} {\bibfnamefont {A.}~\bibnamefont {Megrant}}, \bibinfo {author}
  {\bibfnamefont {E.}~\bibnamefont {Jeffrey}}, \bibinfo {author} {\bibfnamefont
  {T.}~\bibnamefont {White}}, \bibinfo {author} {\bibfnamefont
  {D.}~\bibnamefont {Sank}}, \bibinfo {author} {\bibfnamefont {J.}~\bibnamefont
  {Mutus}}, \bibinfo {author} {\bibfnamefont {B.}~\bibnamefont {Campbell}},
  \bibinfo {author} {\bibfnamefont {Y.}~\bibnamefont {Chen}},  \emph {et~al.},\
  }\href@noop {} {\bibfield  {journal} {\bibinfo  {journal} {Nature}\ }\textbf
  {\bibinfo {volume} {519}},\ \bibinfo {pages} {66} (\bibinfo {year}
  {2015})}\BibitemShut {NoStop}%
\bibitem [{\citenamefont {Corcoles}\ \emph {et~al.}(2015)\citenamefont
  {Corcoles}, \citenamefont {Magesan}, \citenamefont {Srinivasan},
  \citenamefont {Cross}, \citenamefont {Steffen}, \citenamefont {Gambetta},\
  and\ \citenamefont {Chow}}]{Corcoles15}%
  \BibitemOpen
  \bibfield  {author} {\bibinfo {author} {\bibfnamefont {A.~D.}\ \bibnamefont
  {Corcoles}}, \bibinfo {author} {\bibfnamefont {E.}~\bibnamefont {Magesan}},
  \bibinfo {author} {\bibfnamefont {S.~J.}\ \bibnamefont {Srinivasan}},
  \bibinfo {author} {\bibfnamefont {A.~W.}\ \bibnamefont {Cross}}, \bibinfo
  {author} {\bibfnamefont {M.}~\bibnamefont {Steffen}}, \bibinfo {author}
  {\bibfnamefont {J.~M.}\ \bibnamefont {Gambetta}}, \ and\ \bibinfo {author}
  {\bibfnamefont {J.~M.}\ \bibnamefont {Chow}},\ }\href@noop {} {\bibfield
  {journal} {\bibinfo  {journal} {Nat.\ Commun.}\ }\textbf {\bibinfo {volume}
  {{6}}},\ \bibinfo {pages} {6979} (\bibinfo {year} {{2015}})}\BibitemShut
  {NoStop}%
\bibitem [{\citenamefont {Rist\`{e}}\ \emph {et~al.}(2015)\citenamefont
  {Rist\`{e}}, \citenamefont {Poletto}, \citenamefont {Huang}, \citenamefont
  {Bruno}, \citenamefont {Vesterinen}, \citenamefont {Saira},\ and\
  \citenamefont {DiCarlo}}]{Riste15}%
  \BibitemOpen
  \bibfield  {author} {\bibinfo {author} {\bibfnamefont {D.}~\bibnamefont
  {Rist\`{e}}}, \bibinfo {author} {\bibfnamefont {S.}~\bibnamefont {Poletto}},
  \bibinfo {author} {\bibfnamefont {M.~Z.}\ \bibnamefont {Huang}}, \bibinfo
  {author} {\bibfnamefont {A.}~\bibnamefont {Bruno}}, \bibinfo {author}
  {\bibfnamefont {V.}~\bibnamefont {Vesterinen}}, \bibinfo {author}
  {\bibfnamefont {O.~P.}\ \bibnamefont {Saira}}, \ and\ \bibinfo {author}
  {\bibfnamefont {L.}~\bibnamefont {DiCarlo}},\ }\href@noop {} {\bibfield
  {journal} {\bibinfo  {journal} {Nat.\ Commun.}\ }\textbf {\bibinfo {volume}
  {{6}}} (\bibinfo {year} {{2015}})}\BibitemShut {NoStop}%
\bibitem [{\citenamefont {Debnath}\ \emph {et~al.}(2016)\citenamefont
  {Debnath}, \citenamefont {Linke}, \citenamefont {Figgatt}, \citenamefont
  {Landsman}, \citenamefont {Wright},\ and\ \citenamefont
  {Monroe}}]{Debnath16}%
  \BibitemOpen
  \bibfield  {author} {\bibinfo {author} {\bibfnamefont {S.}~\bibnamefont
  {Debnath}}, \bibinfo {author} {\bibfnamefont {N.~M.}\ \bibnamefont {Linke}},
  \bibinfo {author} {\bibfnamefont {C.}~\bibnamefont {Figgatt}}, \bibinfo
  {author} {\bibfnamefont {K.~A.}\ \bibnamefont {Landsman}}, \bibinfo {author}
  {\bibfnamefont {K.}~\bibnamefont {Wright}}, \ and\ \bibinfo {author}
  {\bibfnamefont {C.}~\bibnamefont {Monroe}},\ }\href@noop {} {\bibfield
  {journal} {\bibinfo  {journal} {Nature}\ }\textbf {\bibinfo {volume} {536}},\
  \bibinfo {pages} {63} (\bibinfo {year} {2016})}\BibitemShut {NoStop}%
\bibitem [{\citenamefont {Cramer}\ \emph {et~al.}(2016)\citenamefont {Cramer},
  \citenamefont {Kalb}, \citenamefont {Rol}, \citenamefont {Hensen},
  \citenamefont {Markham}, \citenamefont {Twitchen}, \citenamefont {Hanson},\
  and\ \citenamefont {Taminiau}}]{Cramer16}%
  \BibitemOpen
  \bibfield  {author} {\bibinfo {author} {\bibfnamefont {J.}~\bibnamefont
  {Cramer}}, \bibinfo {author} {\bibfnamefont {N.}~\bibnamefont {Kalb}},
  \bibinfo {author} {\bibfnamefont {M.~A.}\ \bibnamefont {Rol}}, \bibinfo
  {author} {\bibfnamefont {B.}~\bibnamefont {Hensen}}, \bibinfo {author}
  {\bibfnamefont {M.}~\bibnamefont {Markham}}, \bibinfo {author} {\bibfnamefont
  {D.~J.}\ \bibnamefont {Twitchen}}, \bibinfo {author} {\bibfnamefont
  {R.}~\bibnamefont {Hanson}}, \ and\ \bibinfo {author} {\bibfnamefont {T.~H.}\
  \bibnamefont {Taminiau}},\ }\href@noop {} {\bibfield  {journal} {\bibinfo
  {journal} {Nat.\ Commun.}\ }\textbf {\bibinfo {volume} {{5}}},\ \bibinfo
  {pages} {11526} (\bibinfo {year} {{2016}})}\BibitemShut {NoStop}%
\bibitem [{\citenamefont {Nielsen}\ and\ \citenamefont
  {Chuang}(2000)}]{Nielsen00}%
  \BibitemOpen
  \bibfield  {author} {\bibinfo {author} {\bibfnamefont {M.~A.}\ \bibnamefont
  {Nielsen}}\ and\ \bibinfo {author} {\bibfnamefont {I.~L.}\ \bibnamefont
  {Chuang}},\ }\href@noop {} {\emph {\bibinfo {title} {Quantum Computation and
  Quantum Information}}}\ (\bibinfo  {publisher} {Cambridge University Press},\
  \bibinfo {address} {Cambridge},\ \bibinfo {year} {2000})\BibitemShut
  {NoStop}%
\bibitem [{\citenamefont {Martinis}(2016)}]{Martinis16}%
  \BibitemOpen
  \bibfield  {author} {\bibinfo {author} {\bibfnamefont {J.~M.}\ \bibnamefont
  {Martinis}},\ }\href@noop {} {\bibfield  {journal} {\bibinfo  {journal} {npj
  Quantum Inf.}\ }\textbf {\bibinfo {volume} {1}},\ \bibinfo {pages} {15005}
  (\bibinfo {year} {2016})}\BibitemShut {NoStop}%
\bibitem [{\citenamefont {Brown}\ \emph {et~al.}(2016)\citenamefont {Brown},
  \citenamefont {Kim},\ and\ \citenamefont {Monroe}}]{Brown16}%
  \BibitemOpen
  \bibfield  {author} {\bibinfo {author} {\bibfnamefont {K.~R.}\ \bibnamefont
  {Brown}}, \bibinfo {author} {\bibfnamefont {J.}~\bibnamefont {Kim}}, \ and\
  \bibinfo {author} {\bibfnamefont {C.}~\bibnamefont {Monroe}},\ }\href@noop {}
  {\bibfield  {journal} {\bibinfo  {journal} {npj Quantum Inf.}\ }\textbf
  {\bibinfo {volume} {2}},\ \bibinfo {pages} {16034} (\bibinfo {year}
  {2016})}\BibitemShut {NoStop}%
\bibitem [{\citenamefont {Nickerson}\ \emph {et~al.}(2014)\citenamefont
  {Nickerson}, \citenamefont {Fitzsimons},\ and\ \citenamefont
  {Benjamin}}]{Nickerson14}%
  \BibitemOpen
  \bibfield  {author} {\bibinfo {author} {\bibfnamefont {N.~H.}\ \bibnamefont
  {Nickerson}}, \bibinfo {author} {\bibfnamefont {J.~F.}\ \bibnamefont
  {Fitzsimons}}, \ and\ \bibinfo {author} {\bibfnamefont {S.~C.}\ \bibnamefont
  {Benjamin}},\ }\href@noop {} {\bibfield  {journal} {\bibinfo  {journal}
  {Phys. Rev. X}\ }\textbf {\bibinfo {volume} {4}},\ \bibinfo {pages} {041041}
  (\bibinfo {year} {2014})}\BibitemShut {NoStop}%
\bibitem [{\citenamefont {Fowler}\ \emph {et~al.}(2012)\citenamefont {Fowler},
  \citenamefont {Mariantoni}, \citenamefont {Martinis},\ and\ \citenamefont
  {Cleland}}]{Fowler12}%
  \BibitemOpen
  \bibfield  {author} {\bibinfo {author} {\bibfnamefont {A.~G.}\ \bibnamefont
  {Fowler}}, \bibinfo {author} {\bibfnamefont {M.}~\bibnamefont {Mariantoni}},
  \bibinfo {author} {\bibfnamefont {J.~M.}\ \bibnamefont {Martinis}}, \ and\
  \bibinfo {author} {\bibfnamefont {A.~N.}\ \bibnamefont {Cleland}},\
  }\href@noop {} {\bibfield  {journal} {\bibinfo  {journal} {Phys. Rev. A}\
  }\textbf {\bibinfo {volume} {86}},\ \bibinfo {pages} {032324} (\bibinfo
  {year} {2012})}\BibitemShut {NoStop}%
\bibitem [{\citenamefont {Jones}\ \emph {et~al.}(2012)\citenamefont {Jones},
  \citenamefont {Van~Meter}, \citenamefont {Fowler}, \citenamefont {McMahon},
  \citenamefont {Kim}, \citenamefont {Ladd},\ and\ \citenamefont
  {Yamamoto}}]{Jones12}%
  \BibitemOpen
  \bibfield  {author} {\bibinfo {author} {\bibfnamefont {N.~C.}\ \bibnamefont
  {Jones}}, \bibinfo {author} {\bibfnamefont {R.}~\bibnamefont {Van~Meter}},
  \bibinfo {author} {\bibfnamefont {A.~G.}\ \bibnamefont {Fowler}}, \bibinfo
  {author} {\bibfnamefont {P.~L.}\ \bibnamefont {McMahon}}, \bibinfo {author}
  {\bibfnamefont {J.}~\bibnamefont {Kim}}, \bibinfo {author} {\bibfnamefont
  {T.~D.}\ \bibnamefont {Ladd}}, \ and\ \bibinfo {author} {\bibfnamefont
  {Y.}~\bibnamefont {Yamamoto}},\ }\href@noop {} {\bibfield  {journal}
  {\bibinfo  {journal} {Phys. Rev. X}\ }\textbf {\bibinfo {volume} {2}},\
  \bibinfo {pages} {031007} (\bibinfo {year} {2012})}\BibitemShut {NoStop}%
\bibitem [{\citenamefont {Bravyi}\ and\ \citenamefont
  {Kitaev}(1998)}]{Bravyi98}%
  \BibitemOpen
  \bibfield  {author} {\bibinfo {author} {\bibfnamefont {S.~B.}\ \bibnamefont
  {Bravyi}}\ and\ \bibinfo {author} {\bibfnamefont {A.~Y.}\ \bibnamefont
  {Kitaev}},\ }\href@noop {} {\bibfield  {journal} {\bibinfo  {journal}
  {arXiv:quant-ph/9811052}\ } (\bibinfo {year} {1998})}\BibitemShut {NoStop}%
\bibitem [{\citenamefont {Terhal}(2015)}]{Terhal15}%
  \BibitemOpen
  \bibfield  {author} {\bibinfo {author} {\bibfnamefont {B.~M.}\ \bibnamefont
  {Terhal}},\ }\href {\doibase 10.1103/RevModPhys.87.307} {\bibfield  {journal}
  {\bibinfo  {journal} {Rev. Mod. Phys.}\ }\textbf {\bibinfo {volume} {87}},\
  \bibinfo {pages} {307} (\bibinfo {year} {2015})}\BibitemShut {NoStop}%
\bibitem [{\citenamefont {Blais}\ \emph {et~al.}(2004)\citenamefont {Blais},
  \citenamefont {Huang}, \citenamefont {Wallraff}, \citenamefont {Girvin},\
  and\ \citenamefont {Schoelkopf}}]{Blais04}%
  \BibitemOpen
  \bibfield  {author} {\bibinfo {author} {\bibfnamefont {A.}~\bibnamefont
  {Blais}}, \bibinfo {author} {\bibfnamefont {R.-S.}\ \bibnamefont {Huang}},
  \bibinfo {author} {\bibfnamefont {A.}~\bibnamefont {Wallraff}}, \bibinfo
  {author} {\bibfnamefont {S.~M.}\ \bibnamefont {Girvin}}, \ and\ \bibinfo
  {author} {\bibfnamefont {R.~J.}\ \bibnamefont {Schoelkopf}},\ }\href@noop {}
  {\bibfield  {journal} {\bibinfo  {journal} {Phys. Rev. A}\ }\textbf {\bibinfo
  {volume} {69}},\ \bibinfo {pages} {062320} (\bibinfo {year}
  {2004})}\BibitemShut {NoStop}%
\bibitem [{\citenamefont {Barends}\ \emph {et~al.}(2014)\citenamefont
  {Barends}, \citenamefont {Kelly}, \citenamefont {Megrant}, \citenamefont
  {Veitia}, \citenamefont {Sank}, \citenamefont {Jeffrey}, \citenamefont
  {White}, \citenamefont {Mutus}, \citenamefont {Fowler}, \citenamefont
  {Campbell}, \citenamefont {Chen}, \citenamefont {Chen}, \citenamefont
  {Chiaro}, \citenamefont {Dunsworth}, \citenamefont {Neill}, \citenamefont
  {O'Malley}, \citenamefont {Roushan}, \citenamefont {Vainsencher},
  \citenamefont {Wenner}, \citenamefont {Korotkov}, \citenamefont {Cleland},\
  and\ \citenamefont {Martinis}}]{Barends14}%
  \BibitemOpen
  \bibfield  {author} {\bibinfo {author} {\bibfnamefont {R.}~\bibnamefont
  {Barends}}, \bibinfo {author} {\bibfnamefont {J.}~\bibnamefont {Kelly}},
  \bibinfo {author} {\bibfnamefont {A.}~\bibnamefont {Megrant}}, \bibinfo
  {author} {\bibfnamefont {A.}~\bibnamefont {Veitia}}, \bibinfo {author}
  {\bibfnamefont {D.}~\bibnamefont {Sank}}, \bibinfo {author} {\bibfnamefont
  {E.}~\bibnamefont {Jeffrey}}, \bibinfo {author} {\bibfnamefont {T.~C.}\
  \bibnamefont {White}}, \bibinfo {author} {\bibfnamefont {J.}~\bibnamefont
  {Mutus}}, \bibinfo {author} {\bibfnamefont {A.~G.}\ \bibnamefont {Fowler}},
  \bibinfo {author} {\bibfnamefont {B.}~\bibnamefont {Campbell}}, \bibinfo
  {author} {\bibfnamefont {Y.}~\bibnamefont {Chen}}, \bibinfo {author}
  {\bibfnamefont {Z.}~\bibnamefont {Chen}}, \bibinfo {author} {\bibfnamefont
  {B.}~\bibnamefont {Chiaro}}, \bibinfo {author} {\bibfnamefont
  {A.}~\bibnamefont {Dunsworth}}, \bibinfo {author} {\bibfnamefont
  {C.}~\bibnamefont {Neill}}, \bibinfo {author} {\bibfnamefont
  {P.}~\bibnamefont {O'Malley}}, \bibinfo {author} {\bibfnamefont
  {P.}~\bibnamefont {Roushan}}, \bibinfo {author} {\bibfnamefont
  {A.}~\bibnamefont {Vainsencher}}, \bibinfo {author} {\bibfnamefont
  {J.}~\bibnamefont {Wenner}}, \bibinfo {author} {\bibfnamefont {A.~N.}\
  \bibnamefont {Korotkov}}, \bibinfo {author} {\bibfnamefont {A.~N.}\
  \bibnamefont {Cleland}}, \ and\ \bibinfo {author} {\bibfnamefont {J.~M.}\
  \bibnamefont {Martinis}},\ }\href@noop {} {\bibfield  {journal} {\bibinfo
  {journal} {Nature}\ }\textbf {\bibinfo {volume} {508}},\ \bibinfo {pages}
  {500} (\bibinfo {year} {2014})}\BibitemShut {NoStop}%
\bibitem [{\citenamefont {Rol}\ \emph {et~al.}(2016)\citenamefont {Rol},
  \citenamefont {Bultink}, \citenamefont {O'Brien}, \citenamefont {de~Jong},
  \citenamefont {Theis}, \citenamefont {Fu}, \citenamefont {Luthi},
  \citenamefont {Vermeulen}, \citenamefont {de~Sterke}, \citenamefont {Bruno},
  \citenamefont {Deurloo}, \citenamefont {Schouten}, \citenamefont {Wilhelm},\
  and\ \citenamefont {DiCarlo}}]{Rol16}%
  \BibitemOpen
  \bibfield  {author} {\bibinfo {author} {\bibfnamefont {M.~A.}\ \bibnamefont
  {Rol}}, \bibinfo {author} {\bibfnamefont {C.~C.}\ \bibnamefont {Bultink}},
  \bibinfo {author} {\bibfnamefont {T.~E.}\ \bibnamefont {O'Brien}}, \bibinfo
  {author} {\bibfnamefont {S.~R.}\ \bibnamefont {de~Jong}}, \bibinfo {author}
  {\bibfnamefont {L.~S.}\ \bibnamefont {Theis}}, \bibinfo {author}
  {\bibfnamefont {X.}~\bibnamefont {Fu}}, \bibinfo {author} {\bibfnamefont
  {F.}~\bibnamefont {Luthi}}, \bibinfo {author} {\bibfnamefont {R.~F.~L.}\
  \bibnamefont {Vermeulen}}, \bibinfo {author} {\bibfnamefont {J.~C.}\
  \bibnamefont {de~Sterke}}, \bibinfo {author} {\bibfnamefont {A.}~\bibnamefont
  {Bruno}}, \bibinfo {author} {\bibfnamefont {D.}~\bibnamefont {Deurloo}},
  \bibinfo {author} {\bibfnamefont {R.~N.}\ \bibnamefont {Schouten}}, \bibinfo
  {author} {\bibfnamefont {F.~K.}\ \bibnamefont {Wilhelm}}, \ and\ \bibinfo
  {author} {\bibfnamefont {L.}~\bibnamefont {DiCarlo}},\ }\href@noop {}
  {\bibfield  {journal} {\bibinfo  {journal} {ArVix:1611.04815}\ } (\bibinfo
  {year} {2016})}\BibitemShut {NoStop}%
\bibitem [{\citenamefont {Rist\`e}\ \emph {et~al.}(2012)\citenamefont
  {Rist\`e}, \citenamefont {Bultink}, \citenamefont {Lehnert},\ and\
  \citenamefont {DiCarlo}}]{Riste12b}%
  \BibitemOpen
  \bibfield  {author} {\bibinfo {author} {\bibfnamefont {D.}~\bibnamefont
  {Rist\`e}}, \bibinfo {author} {\bibfnamefont {C.~C.}\ \bibnamefont
  {Bultink}}, \bibinfo {author} {\bibfnamefont {K.~W.}\ \bibnamefont
  {Lehnert}}, \ and\ \bibinfo {author} {\bibfnamefont {L.}~\bibnamefont
  {DiCarlo}},\ }\href@noop {} {\bibfield  {journal} {\bibinfo  {journal} {Phys.
  Rev. Lett.}\ }\textbf {\bibinfo {volume} {109}},\ \bibinfo {pages} {240502}
  (\bibinfo {year} {2012})}\BibitemShut {NoStop}%
\bibitem [{\citenamefont {Jeffrey}\ \emph {et~al.}(2014)\citenamefont
  {Jeffrey}, \citenamefont {Sank}, \citenamefont {Mutus}, \citenamefont
  {White}, \citenamefont {Kelly}, \citenamefont {Barends}, \citenamefont
  {Chen}, \citenamefont {Chen}, \citenamefont {Chiaro}, \citenamefont
  {Dunsworth}, \citenamefont {Megrant}, \citenamefont {O'Malley}, \citenamefont
  {Neill}, \citenamefont {Roushan}, \citenamefont {Vainsencher}, \citenamefont
  {Wenner}, \citenamefont {Cleland},\ and\ \citenamefont
  {Martinis}}]{Jeffrey14}%
  \BibitemOpen
  \bibfield  {author} {\bibinfo {author} {\bibfnamefont {E.}~\bibnamefont
  {Jeffrey}}, \bibinfo {author} {\bibfnamefont {D.}~\bibnamefont {Sank}},
  \bibinfo {author} {\bibfnamefont {J.~Y.}\ \bibnamefont {Mutus}}, \bibinfo
  {author} {\bibfnamefont {T.~C.}\ \bibnamefont {White}}, \bibinfo {author}
  {\bibfnamefont {J.}~\bibnamefont {Kelly}}, \bibinfo {author} {\bibfnamefont
  {R.}~\bibnamefont {Barends}}, \bibinfo {author} {\bibfnamefont
  {Y.}~\bibnamefont {Chen}}, \bibinfo {author} {\bibfnamefont {Z.}~\bibnamefont
  {Chen}}, \bibinfo {author} {\bibfnamefont {B.}~\bibnamefont {Chiaro}},
  \bibinfo {author} {\bibfnamefont {A.}~\bibnamefont {Dunsworth}}, \bibinfo
  {author} {\bibfnamefont {A.}~\bibnamefont {Megrant}}, \bibinfo {author}
  {\bibfnamefont {P.~J.~J.}\ \bibnamefont {O'Malley}}, \bibinfo {author}
  {\bibfnamefont {C.}~\bibnamefont {Neill}}, \bibinfo {author} {\bibfnamefont
  {P.}~\bibnamefont {Roushan}}, \bibinfo {author} {\bibfnamefont
  {A.}~\bibnamefont {Vainsencher}}, \bibinfo {author} {\bibfnamefont
  {J.}~\bibnamefont {Wenner}}, \bibinfo {author} {\bibfnamefont {A.~N.}\
  \bibnamefont {Cleland}}, \ and\ \bibinfo {author} {\bibfnamefont {J.~M.}\
  \bibnamefont {Martinis}},\ }\href@noop {} {\bibfield  {journal} {\bibinfo
  {journal} {Phys. Rev. Lett.}\ }\textbf {\bibinfo {volume} {112}},\ \bibinfo
  {pages} {190504} (\bibinfo {year} {2014})}\BibitemShut {NoStop}%
\bibitem [{\citenamefont {Bruno}\ \emph {et~al.}()\citenamefont {Bruno},
  \citenamefont {Poletto}, \citenamefont {Haider},\ and\ \citenamefont
  {DiCarlo}}]{Bruno16MM}%
  \BibitemOpen
  \bibfield  {author} {\bibinfo {author} {\bibfnamefont {A.}~\bibnamefont
  {Bruno}}, \bibinfo {author} {\bibfnamefont {S.}~\bibnamefont {Poletto}},
  \bibinfo {author} {\bibfnamefont {N.}~\bibnamefont {Haider}}, \ and\ \bibinfo
  {author} {\bibfnamefont {L.}~\bibnamefont {DiCarlo}},\ }\href@noop {}
  {\enquote {\bibinfo {title} {X48.00004: Extensible circuit {QED} processor
  architecture with vertical {I/O}},}\ }\bibinfo {howpublished} {APS March
  Meeting 2016}\BibitemShut {NoStop}%
\bibitem [{\citenamefont {Bejanin}\ \emph {et~al.}(2016)\citenamefont
  {Bejanin}, \citenamefont {McConkey}, \citenamefont {Rinehart}, \citenamefont
  {Earnest}, \citenamefont {McRae}, \citenamefont {Shiri}, \citenamefont
  {Bateman}, \citenamefont {Rohanizadegan}, \citenamefont {Penava},
  \citenamefont {Breul}, \citenamefont {Royak}, \citenamefont {Zapatka},
  \citenamefont {Fowler},\ and\ \citenamefont {Mariantoni}}]{Bejanin16}%
  \BibitemOpen
  \bibfield  {author} {\bibinfo {author} {\bibfnamefont {J.~H.}\ \bibnamefont
  {Bejanin}}, \bibinfo {author} {\bibfnamefont {T.~G.}\ \bibnamefont
  {McConkey}}, \bibinfo {author} {\bibfnamefont {J.~R.}\ \bibnamefont
  {Rinehart}}, \bibinfo {author} {\bibfnamefont {C.~T.}\ \bibnamefont
  {Earnest}}, \bibinfo {author} {\bibfnamefont {C.~R.~H.}\ \bibnamefont
  {McRae}}, \bibinfo {author} {\bibfnamefont {D.}~\bibnamefont {Shiri}},
  \bibinfo {author} {\bibfnamefont {J.~D.}\ \bibnamefont {Bateman}}, \bibinfo
  {author} {\bibfnamefont {Y.}~\bibnamefont {Rohanizadegan}}, \bibinfo {author}
  {\bibfnamefont {B.}~\bibnamefont {Penava}}, \bibinfo {author} {\bibfnamefont
  {P.}~\bibnamefont {Breul}}, \bibinfo {author} {\bibfnamefont
  {S.}~\bibnamefont {Royak}}, \bibinfo {author} {\bibfnamefont
  {M.}~\bibnamefont {Zapatka}}, \bibinfo {author} {\bibfnamefont {A.~G.}\
  \bibnamefont {Fowler}}, \ and\ \bibinfo {author} {\bibfnamefont
  {M.}~\bibnamefont {Mariantoni}},\ }\href@noop {} {\bibfield  {journal}
  {\bibinfo  {journal} {ArXiv:1606.00063}\ } (\bibinfo {year}
  {2016})}\BibitemShut {NoStop}%
\bibitem [{\citenamefont {Rosenberg}\ \emph {et~al.}()\citenamefont
  {Rosenberg}, \citenamefont {Yost}, \citenamefont {Das}, \citenamefont
  {Hover}, \citenamefont {Racz}, \citenamefont {Weber}, \citenamefont {Yoder},
  \citenamefont {Kerman},\ and\ \citenamefont {Oliver}}]{Rosenberg16MM}%
  \BibitemOpen
  \bibfield  {author} {\bibinfo {author} {\bibfnamefont {D.}~\bibnamefont
  {Rosenberg}}, \bibinfo {author} {\bibfnamefont {D.~R.}\ \bibnamefont {Yost}},
  \bibinfo {author} {\bibfnamefont {R.}~\bibnamefont {Das}}, \bibinfo {author}
  {\bibfnamefont {D.}~\bibnamefont {Hover}}, \bibinfo {author} {\bibfnamefont
  {L.}~\bibnamefont {Racz}}, \bibinfo {author} {\bibfnamefont {S.}~\bibnamefont
  {Weber}}, \bibinfo {author} {\bibfnamefont {J.}~\bibnamefont {Yoder}},
  \bibinfo {author} {\bibfnamefont {A.}~\bibnamefont {Kerman}}, \ and\ \bibinfo
  {author} {\bibfnamefont {W.~D.}\ \bibnamefont {Oliver}},\ }\href@noop {}
  {\enquote {\bibinfo {title} {X48.00003: {3D} integration for superconducting
  qubits},}\ }\bibinfo {howpublished} {APS March Meeting 2016}\BibitemShut
  {NoStop}%
\bibitem [{\citenamefont {Groen}\ \emph {et~al.}(2013)\citenamefont {Groen},
  \citenamefont {Rist\`e}, \citenamefont {Tornberg}, \citenamefont {Cramer},
  \citenamefont {de~Groot}, \citenamefont {Picot}, \citenamefont {Johansson},\
  and\ \citenamefont {DiCarlo}}]{Groen13}%
  \BibitemOpen
  \bibfield  {author} {\bibinfo {author} {\bibfnamefont {J.~P.}\ \bibnamefont
  {Groen}}, \bibinfo {author} {\bibfnamefont {D.}~\bibnamefont {Rist\`e}},
  \bibinfo {author} {\bibfnamefont {L.}~\bibnamefont {Tornberg}}, \bibinfo
  {author} {\bibfnamefont {J.}~\bibnamefont {Cramer}}, \bibinfo {author}
  {\bibfnamefont {P.~C.}\ \bibnamefont {de~Groot}}, \bibinfo {author}
  {\bibfnamefont {T.}~\bibnamefont {Picot}}, \bibinfo {author} {\bibfnamefont
  {G.}~\bibnamefont {Johansson}}, \ and\ \bibinfo {author} {\bibfnamefont
  {L.}~\bibnamefont {DiCarlo}},\ }\href@noop {} {\bibfield  {journal} {\bibinfo
   {journal} {Phys. Rev. Lett.}\ }\textbf {\bibinfo {volume} {111}},\ \bibinfo
  {pages} {090506} (\bibinfo {year} {2013})}\BibitemShut {NoStop}%
\bibitem [{\citenamefont {Asaad}\ \emph {et~al.}(2016)\citenamefont {Asaad},
  \citenamefont {Dickel}, \citenamefont {Poletto}, \citenamefont {Bruno},
  \citenamefont {Langford}, \citenamefont {Rol}, \citenamefont {Deurloo},\ and\
  \citenamefont {DiCarlo}}]{Asaad16}%
  \BibitemOpen
  \bibfield  {author} {\bibinfo {author} {\bibfnamefont {S.}~\bibnamefont
  {Asaad}}, \bibinfo {author} {\bibfnamefont {C.}~\bibnamefont {Dickel}},
  \bibinfo {author} {\bibfnamefont {S.}~\bibnamefont {Poletto}}, \bibinfo
  {author} {\bibfnamefont {A.}~\bibnamefont {Bruno}}, \bibinfo {author}
  {\bibfnamefont {N.~K.}\ \bibnamefont {Langford}}, \bibinfo {author}
  {\bibfnamefont {M.~A.}\ \bibnamefont {Rol}}, \bibinfo {author} {\bibfnamefont
  {D.}~\bibnamefont {Deurloo}}, \ and\ \bibinfo {author} {\bibfnamefont
  {L.}~\bibnamefont {DiCarlo}},\ }\href@noop {} {\bibfield  {journal} {\bibinfo
   {journal} {npj Quantum Inf.}\ }\textbf {\bibinfo {volume} {2}},\ \bibinfo
  {pages} {16029} (\bibinfo {year} {2016})}\BibitemShut {NoStop}%
\bibitem [{\citenamefont {Strauch}\ \emph {et~al.}(2003)\citenamefont
  {Strauch}, \citenamefont {Johnson}, \citenamefont {Dragt}, \citenamefont
  {Lobb}, \citenamefont {Anderson},\ and\ \citenamefont
  {Wellstood}}]{Strauch03}%
  \BibitemOpen
  \bibfield  {author} {\bibinfo {author} {\bibfnamefont {F.~W.}\ \bibnamefont
  {Strauch}}, \bibinfo {author} {\bibfnamefont {P.~R.}\ \bibnamefont
  {Johnson}}, \bibinfo {author} {\bibfnamefont {A.~J.}\ \bibnamefont {Dragt}},
  \bibinfo {author} {\bibfnamefont {C.~J.}\ \bibnamefont {Lobb}}, \bibinfo
  {author} {\bibfnamefont {J.~R.}\ \bibnamefont {Anderson}}, \ and\ \bibinfo
  {author} {\bibfnamefont {F.~C.}\ \bibnamefont {Wellstood}},\ }\href@noop {}
  {\bibfield  {journal} {\bibinfo  {journal} {Phys. Rev. Lett.}\ }\textbf
  {\bibinfo {volume} {91}},\ \bibinfo {pages} {167005} (\bibinfo {year}
  {2003})}\BibitemShut {NoStop}%
\bibitem [{\citenamefont {Di{C}arlo}\ \emph {et~al.}(2009)\citenamefont
  {Di{C}arlo}, \citenamefont {Chow}, \citenamefont {Gambetta}, \citenamefont
  {Bishop}, \citenamefont {Johnson}, \citenamefont {Schuster}, \citenamefont
  {Majer}, \citenamefont {Blas}, \citenamefont {Frunzio},\ and\ \citenamefont
  {Schoelkopf}}]{DiCarlo09}%
  \BibitemOpen
  \bibfield  {author} {\bibinfo {author} {\bibfnamefont {L.}~\bibnamefont
  {Di{C}arlo}}, \bibinfo {author} {\bibfnamefont {J.~M.}\ \bibnamefont {Chow}},
  \bibinfo {author} {\bibfnamefont {J.~M.}\ \bibnamefont {Gambetta}}, \bibinfo
  {author} {\bibfnamefont {L.~S.}\ \bibnamefont {Bishop}}, \bibinfo {author}
  {\bibfnamefont {B.~R.}\ \bibnamefont {Johnson}}, \bibinfo {author}
  {\bibfnamefont {D.~I.}\ \bibnamefont {Schuster}}, \bibinfo {author}
  {\bibfnamefont {J.}~\bibnamefont {Majer}}, \bibinfo {author} {\bibfnamefont
  {A.}~\bibnamefont {Blas}}, \bibinfo {author} {\bibfnamefont {S.~M.}\
  \bibnamefont {Frunzio}, \bibfnamefont {L.~an~Girvin}}, \ and\ \bibinfo
  {author} {\bibfnamefont {R.~J.}\ \bibnamefont {Schoelkopf}},\ }\href
  {http://www.nature.com/nature/journal/v460/n7252/abs/nature08121.html}
  {\bibfield  {journal} {\bibinfo  {journal} {Nature}\ }\textbf {\bibinfo
  {volume} {460}},\ \bibinfo {pages} {240} (\bibinfo {year}
  {2009})}\BibitemShut {NoStop}%
\bibitem [{\citenamefont {DiCarlo}\ \emph {et~al.}(2010)\citenamefont
  {DiCarlo}, \citenamefont {Reed}, \citenamefont {Sun}, \citenamefont
  {Johnson}, \citenamefont {Chow}, \citenamefont {Gambetta}, \citenamefont
  {Frunzio}, \citenamefont {Girvin}, \citenamefont {Devoret},\ and\
  \citenamefont {Schoelkopf}}]{DiCarlo10}%
  \BibitemOpen
  \bibfield  {author} {\bibinfo {author} {\bibfnamefont {L.}~\bibnamefont
  {DiCarlo}}, \bibinfo {author} {\bibfnamefont {M.~D.}\ \bibnamefont {Reed}},
  \bibinfo {author} {\bibfnamefont {L.}~\bibnamefont {Sun}}, \bibinfo {author}
  {\bibfnamefont {B.~R.}\ \bibnamefont {Johnson}}, \bibinfo {author}
  {\bibfnamefont {J.~M.}\ \bibnamefont {Chow}}, \bibinfo {author}
  {\bibfnamefont {J.~M.}\ \bibnamefont {Gambetta}}, \bibinfo {author}
  {\bibfnamefont {L.}~\bibnamefont {Frunzio}}, \bibinfo {author} {\bibfnamefont
  {S.~M.}\ \bibnamefont {Girvin}}, \bibinfo {author} {\bibfnamefont {M.~H.}\
  \bibnamefont {Devoret}}, \ and\ \bibinfo {author} {\bibfnamefont {R.~J.}\
  \bibnamefont {Schoelkopf}},\ }\href@noop {} {\bibfield  {journal} {\bibinfo
  {journal} {Nature}\ }\textbf {\bibinfo {volume} {467}},\ \bibinfo {pages}
  {574} (\bibinfo {year} {2010})}\BibitemShut {NoStop}%
\bibitem [{\citenamefont {Martinis}(2015)}]{Martinis15}%
  \BibitemOpen
  \bibfield  {author} {\bibinfo {author} {\bibfnamefont {J.~M.}\ \bibnamefont
  {Martinis}},\ }\href@noop {} {\bibfield  {journal} {\bibinfo  {journal} {npj
  Quantum Inf.}\ }\textbf {\bibinfo {volume} {1}},\ \bibinfo {pages} {15005}
  (\bibinfo {year} {2015})}\BibitemShut {NoStop}%
\bibitem [{\citenamefont {Horsman}\ \emph {et~al.}(2012)\citenamefont
  {Horsman}, \citenamefont {Fowler}, \citenamefont {Devitt},\ and\
  \citenamefont {Meter}}]{Horsman12}%
  \BibitemOpen
  \bibfield  {author} {\bibinfo {author} {\bibfnamefont {C.}~\bibnamefont
  {Horsman}}, \bibinfo {author} {\bibfnamefont {A.~G.}\ \bibnamefont {Fowler}},
  \bibinfo {author} {\bibfnamefont {S.}~\bibnamefont {Devitt}}, \ and\ \bibinfo
  {author} {\bibfnamefont {R.~V.}\ \bibnamefont {Meter}},\ }\href@noop {}
  {\bibfield  {journal} {\bibinfo  {journal} {New J.\ Phys.}\ }\textbf
  {\bibinfo {volume} {14}},\ \bibinfo {pages} {123011} (\bibinfo {year}
  {2012})}\BibitemShut {NoStop}%
\bibitem [{\citenamefont {Knill}(2005)}]{Knill05}%
  \BibitemOpen
  \bibfield  {author} {\bibinfo {author} {\bibfnamefont {E.}~\bibnamefont
  {Knill}},\ }\href {\doibase 10.1038/nature03350} {\bibfield  {journal}
  {\bibinfo  {journal} {Nature}\ }\textbf {\bibinfo {volume} {434}},\ \bibinfo
  {pages} {39} (\bibinfo {year} {2005})}\BibitemShut {NoStop}%
\bibitem [{\citenamefont {Tomita}\ and\ \citenamefont
  {Svore}(2014)}]{Tomita14}%
  \BibitemOpen
  \bibfield  {author} {\bibinfo {author} {\bibfnamefont {Y.}~\bibnamefont
  {Tomita}}\ and\ \bibinfo {author} {\bibfnamefont {K.~M.}\ \bibnamefont
  {Svore}},\ }\href@noop {} {\bibfield  {journal} {\bibinfo  {journal} {Phys.
  Rev. A}\ }\textbf {\bibinfo {volume} {90}},\ \bibinfo {pages} {062320}
  (\bibinfo {year} {2014})}\BibitemShut {NoStop}%
\bibitem [{\citenamefont {Schreier}\ \emph {et~al.}(2008)\citenamefont
  {Schreier}, \citenamefont {Houck}, \citenamefont {Koch}, \citenamefont
  {Schuster}, \citenamefont {Johnson}, \citenamefont {Chow}, \citenamefont
  {Gambetta}, \citenamefont {Majer}, \citenamefont {Frunzio}, \citenamefont
  {Devoret}, \citenamefont {Girvin},\ and\ \citenamefont
  {Schoelkopf}}]{Schreier08}%
  \BibitemOpen
  \bibfield  {author} {\bibinfo {author} {\bibfnamefont {J.~A.}\ \bibnamefont
  {Schreier}}, \bibinfo {author} {\bibfnamefont {A.~A.}\ \bibnamefont {Houck}},
  \bibinfo {author} {\bibfnamefont {J.}~\bibnamefont {Koch}}, \bibinfo {author}
  {\bibfnamefont {D.~I.}\ \bibnamefont {Schuster}}, \bibinfo {author}
  {\bibfnamefont {B.~R.}\ \bibnamefont {Johnson}}, \bibinfo {author}
  {\bibfnamefont {J.~M.}\ \bibnamefont {Chow}}, \bibinfo {author}
  {\bibfnamefont {J.~M.}\ \bibnamefont {Gambetta}}, \bibinfo {author}
  {\bibfnamefont {J.}~\bibnamefont {Majer}}, \bibinfo {author} {\bibfnamefont
  {L.}~\bibnamefont {Frunzio}}, \bibinfo {author} {\bibfnamefont {M.~H.}\
  \bibnamefont {Devoret}}, \bibinfo {author} {\bibfnamefont {S.~M.}\
  \bibnamefont {Girvin}}, \ and\ \bibinfo {author} {\bibfnamefont {R.~J.}\
  \bibnamefont {Schoelkopf}},\ }\href@noop {} {\bibfield  {journal} {\bibinfo
  {journal} {Phys. Rev. B}\ }\textbf {\bibinfo {volume} {77}},\ \bibinfo
  {pages} {180502} (\bibinfo {year} {2008})}\BibitemShut {NoStop}%
\bibitem [{\citenamefont {Koch}\ \emph {et~al.}(2007)\citenamefont {Koch},
  \citenamefont {Yu}, \citenamefont {Gambetta}, \citenamefont {Houck},
  \citenamefont {Schuster}, \citenamefont {Majer}, \citenamefont {Blais},
  \citenamefont {Devoret}, \citenamefont {Girvin},\ and\ \citenamefont
  {Schoelkopf}}]{Koch07}%
  \BibitemOpen
  \bibfield  {author} {\bibinfo {author} {\bibfnamefont {J.}~\bibnamefont
  {Koch}}, \bibinfo {author} {\bibfnamefont {T.~M.}\ \bibnamefont {Yu}},
  \bibinfo {author} {\bibfnamefont {J.}~\bibnamefont {Gambetta}}, \bibinfo
  {author} {\bibfnamefont {A.~A.}\ \bibnamefont {Houck}}, \bibinfo {author}
  {\bibfnamefont {D.~I.}\ \bibnamefont {Schuster}}, \bibinfo {author}
  {\bibfnamefont {J.}~\bibnamefont {Majer}}, \bibinfo {author} {\bibfnamefont
  {A.}~\bibnamefont {Blais}}, \bibinfo {author} {\bibfnamefont {M.~H.}\
  \bibnamefont {Devoret}}, \bibinfo {author} {\bibfnamefont {S.~M.}\
  \bibnamefont {Girvin}}, \ and\ \bibinfo {author} {\bibfnamefont {R.~J.}\
  \bibnamefont {Schoelkopf}},\ }\href
  {http://journals.aps.org/pra/abstract/10.1103/PhysRevA.76.042319} {\bibfield
  {journal} {\bibinfo  {journal} {Phys. Rev. A}\ }\textbf {\bibinfo {volume}
  {76}},\ \bibinfo {pages} {042319} (\bibinfo {year} {2007})}\BibitemShut
  {NoStop}%
\bibitem [{\citenamefont {O'Brien\emph{~et~al.}}()}]{OBrien16}%
  \BibitemOpen
  \bibfield  {author} {\bibinfo {author} {\bibfnamefont {T.~E.}\ \bibnamefont
  {O'Brien\emph{~et~al.}}},\ }\href@noop {} {}\bibinfo {howpublished} {in
  preparation}\BibitemShut {NoStop}%
\bibitem [{\citenamefont {Poletto\emph{~et~al.}}()}]{Poletto17}%
  \BibitemOpen
  \bibfield  {author} {\bibinfo {author} {\bibfnamefont {S.}~\bibnamefont
  {Poletto\emph{~et~al.}}},\ }\href@noop {} {}\bibinfo {howpublished} {in
  preparation}\BibitemShut {NoStop}%
\bibitem [{\citenamefont {Motzoi}\ \emph {et~al.}(2009)\citenamefont {Motzoi},
  \citenamefont {Gambetta}, \citenamefont {Rebentrost},\ and\ \citenamefont
  {Wilhelm}}]{Motzoi09}%
  \BibitemOpen
  \bibfield  {author} {\bibinfo {author} {\bibfnamefont {F.}~\bibnamefont
  {Motzoi}}, \bibinfo {author} {\bibfnamefont {J.~M.}\ \bibnamefont
  {Gambetta}}, \bibinfo {author} {\bibfnamefont {P.}~\bibnamefont
  {Rebentrost}}, \ and\ \bibinfo {author} {\bibfnamefont {F.~K.}\ \bibnamefont
  {Wilhelm}},\ }\href@noop {} {\bibfield  {journal} {\bibinfo  {journal} {Phys.
  Rev. Lett.}\ }\textbf {\bibinfo {volume} {103}},\ \bibinfo {pages} {110501}
  (\bibinfo {year} {2009})}\BibitemShut {NoStop}%
\bibitem [{\citenamefont {Chow}\ \emph {et~al.}(2010)\citenamefont {Chow},
  \citenamefont {DiCarlo}, \citenamefont {Gambetta}, \citenamefont {Motzoi},
  \citenamefont {Frunzio}, \citenamefont {Girvin},\ and\ \citenamefont
  {Schoelkopf}}]{Chow10b}%
  \BibitemOpen
  \bibfield  {author} {\bibinfo {author} {\bibfnamefont {J.~M.}\ \bibnamefont
  {Chow}}, \bibinfo {author} {\bibfnamefont {L.}~\bibnamefont {DiCarlo}},
  \bibinfo {author} {\bibfnamefont {J.~M.}\ \bibnamefont {Gambetta}}, \bibinfo
  {author} {\bibfnamefont {F.}~\bibnamefont {Motzoi}}, \bibinfo {author}
  {\bibfnamefont {L.}~\bibnamefont {Frunzio}}, \bibinfo {author} {\bibfnamefont
  {S.~M.}\ \bibnamefont {Girvin}}, \ and\ \bibinfo {author} {\bibfnamefont
  {R.~J.}\ \bibnamefont {Schoelkopf}},\ }\href@noop {} {\bibfield  {journal}
  {\bibinfo  {journal} {Phys. Rev. A}\ }\textbf {\bibinfo {volume} {82}},\
  \bibinfo {pages} {040305} (\bibinfo {year} {2010})}\BibitemShut {NoStop}%
\end{thebibliography}

\end{document}